\documentclass[12pt]{article}
\def\lapprox{\lower .7ex\hbox{$\;\stackrel{\textstyle <}{\sim}\;$}}
\def\gapprox{\lower .7ex\hbox{$\;\stackrel{\textstyle >}{\sim}\;$}}

\def\d{{\rm d}}

\begin{document}

\begin{titlepage}
\vspace*{-1cm}
\begin{flushright}
TTP00--03\\
March 2000 \\
\end{flushright}                                
\vskip 3.5cm
\begin{center}
{\Large\bf Measuring $F_L(x,Q^2)/F_2(x,Q^2)$ from Azimuthal\\[0.2cm]
 Asymmetries
in Deep Inelastic Scattering}
\vskip 1.cm
{\large  T.~Gehrmann}
\vskip .7cm
{\it Institut f\"ur Theoretische Teilchenphysik,
Universit\"at Karlsruhe, D-76128 Karlsruhe, Germany}
\end{center}
\vskip 2.6cm

\begin{abstract}
We demonstrate that the angular
distribution of hadrons produced in semi-inclusive
deep inelastic final states is related to the 
inclusive longitudinal structure function. This relation could provide
a new method of accessing $F_L(x,Q^2)$ in deep inelastic scattering
measurements. 

\end{abstract}

\vfill
\end{titlepage}                                                                
\newpage   

The internal structure of the proton as probed in deep inelastic
lepton-proton scattering can be parameterised in terms of two structure
functions $F_2(x,Q^2)$ and $F_L(x,Q^2)$:
\begin{equation}
\frac{\d \sigma}{\d x \d Q^2} = \frac{2\pi \alpha^2}{xQ^4} 
\left [ \left ( 1+(1-y)^2 \right) F_2(x,Q^2) - y^2 F_L(x,Q^2)\right]\; .
\label{eq:incl}
\end{equation}
$x$ and $Q^2$ are the usual deep inelastic scattering variables related
to the momentum of the outgoing electron, and $y=Q^2/(xs)$, with $s$
being the lepton-proton centre-of-mass energy squared.

While measurements of $F_2(x,Q^2)$ have been made with high precision
over a large kinematical range
(see e.g.~\cite{disrev} for a review), experimental determinations on 
$F_L(x,Q^2)$ are still limited. Given that $F_L(x,Q^2)$, apart from 
encoding information on the internal structure of the proton, plays an 
important role in QED radiative corrections to processes in
lepton-proton scattering, a precise determination of this structure
function over a wide kinematical range would be very desirable.

The experimental measurement of $F_L(x,Q^2)$  in inclusive deep
inelastic scattering is  far from trivial, since the inclusive 
cross section (\ref{eq:incl})
is dominated by the structure function $F_2(x,Q^2)$, with 
$F_L(x,Q^2)$ entering as a correction only. The contributions from
$F_2(x,Q^2)$ and $F_L(x,Q^2)$ to the deep inelastic cross section 
 can be disentangled only from their
kinematical factors; a measurement of $F_L(x,Q^2)$ requires 
therefore the comparison of data at fixed $(x,Q^2)$ taken at different
centre-of-mass energies. Such  measurements have so far been carried
out only at fixed target energies. A determination of $F_L(x,Q^2)$ at
HERA energies has up to now 
been made only in an indirect manner~\cite{flh1} by 
assuming the contribution to the deep inelastic cross section from 
$F_2(x,Q^2)$ to be known from a next-to-leading order QCD fit. 

In this note, we investigate an alternative possibility to determine
$F_L(x,Q^2)$. We argue that the azimuthal distribution of hadrons in
semi-inclusive deep inelastic scattering can be related to the
longitudinal structure function by integrating out the final state
hadron momentum in a appropriate way to recover inclusive expressions. 

The cross section for semi-inclusive hadron production 
reads~\cite{def,men,hag}:
\begin{equation}
\frac{\d \sigma}{\d x\d Q^2 \d z \d \phi} = \frac{\alpha^2 \pi}{2Q^4 z} 
\,\left( A + B \cos \phi + C \cos 2\phi \right) \; .
\label{eq:diff}
\end{equation}
$z$ is the longitudinal
momentum transfer from virtual photon to the final state hadron, and
$\phi$ is the angle between the outgoing lepton direction and outgoing
hadron direction measured in the 
plane transverse to the virtual photon direction. 
The coefficients $A$--$C$ can be computed in the parton model, they
correspond to convolutions of parton distributions $D_{a/N}$ 
in the target, hard subprocess coefficient functions
$f^{b/a}$, and partonic 
fragmentation functions into the final state hadron $D_{h/b}$. 
While $A$ is of ${\cal O}(1)$ in perturbation 
theory, $B$ and $C$ are of ${\cal
O}(\alpha_s)$. For neutral 
current deep inelastic scattering at 
 HERA energies, one finds~\cite{ahmed} that the
coefficients of the angular
dependent terms can be expected to
amount up to 10\% of the constant term, which is
confirmed by a recent measurement~\cite{zeus} by the ZEUS collaboration. 

Momentum conservation in  partonic fragmentation implies 
that the second moment of the fragmentation functions summed over all 
hadrons has to be unity for any parton $b$: 
\begin{equation}
\sum_h \int_0^1 \d z\; z\; D_{h/b} (z,Q^2) = 1 \;.
\end{equation}
This relation is preserved under perturbative evolution. 
Using this relation, it is possible to relate the second moment of any 
semi-inclusive observable to the corresponding inclusive observable. 
The use of this relation in deriving sum rules for semi-inclusive 
observables in deep inelastic scattering has first been suggested 
in~\cite{men}.

Integrating over the angle $\phi$ and 
taking the second moment in $z$ of (\ref{eq:diff}), one recovers the 
inclusive cross section~(\ref{eq:incl}):
\begin{equation}
\sum_h \int_0^1 \d z \int_0^{2\pi} \d \phi \; z \;
\frac{\d \sigma}{\d x\d Q^2 \d z \d \phi} = 
\frac{2\pi \alpha^2}{xQ^4} \left[
\left ( 1+(1-y)^2 \right) F_2(x,Q^2) - y^2 F_L(x,Q^2)\right]\; .
\end{equation}
The coefficient of $\cos 2\phi$ can be expressed by a semi-inclusive
structure function~\cite{hag}:
\begin{equation}
C=(1-y) F_6(x,Q^2,z)\;,
\end{equation}
with
\begin{equation}
F_6(x,Q^2,z) = \frac{z}{\pi}\sum_{a,b} \int_x^1\frac{\d\xi}{\xi} 
\int_z^1\frac{\d \eta}{\eta}\,
D_{h/b}(\eta,Q^2)\, D_{a/N} (\xi,Q^2)\, 4 e_{b/a}^2\, \frac{\alpha_s}{2\pi}
\, f_6^{b/a} \left(\frac{x}{\xi},\frac{z}{\eta}\right) \;,
\end{equation}
where
\begin{eqnarray}
f_6^{q/q}(\hat{x},\hat{z})= f_6^{g/q}(\hat{x},1-\hat{z})
 & =&  2 C_F \hat{x} \hat{z}\; , \nonumber \\
f_6^{q/g}(\hat{x},\hat{z}) = 4T_F \hat{x}(1-\hat{x})\;.
\end{eqnarray} 
In the above $q$ is representing a quark or antiquark. 

Taking the second moment of $f_6^{b/a}(\hat{x},\hat{z})$ with respect to 
$\hat{z}$ and summing over the 
final state parton, one recovers the 
longitudinal coefficient functions:
\begin{eqnarray}
\sum_{b=q,g} \int_0^1 \d \hat z \; \hat{z} f_6^{b/q} (\hat{x},\hat{z})
= C_F \hat x &=& \frac{1}{2}\, C^{(1)}_{L,q}(\hat x) \nonumber \\
\sum_{b=q,\bar q} \int_0^1 \d \hat z \; \hat{z} f_6^{b/g} (\hat{x},\hat{z})
=  4T_F \hat{x}(1-\hat{x}) &=& \frac{1}{2}\, C^{(1)}_{L,g}(\hat x) \;.
\end{eqnarray}

Consequently, one finds that the ratio of the second moment of the 
coefficients of the $\cos 2\phi$ term and the constant term in 
(\ref{eq:diff}) yields:
\begin{eqnarray}
A_{\cos 2\phi}^{(2)}(x,Q^2) &\equiv&
\frac{\displaystyle \sum_h \int_0^1 \d z \int_0^{2\pi}\d \phi 
\; z \; \cos 2\phi \; \frac{\d \sigma}{\d x \d Q^2 \d z \d \phi}}
{\displaystyle \sum_h \int_0^1 \d z \int_0^{2\pi}\d \phi \;
z \; \frac{\d \sigma}{\d x \d Q^2 \d z \d \phi}}\label{eq:master} \\
& =&
  \frac{1}{2}\, \frac{(1-y)F_L(x,Q^2)}{\left(1+(1-y)^2\right)F_2(x,Q^2)-y^2
F_L(x,Q^2)}\;.
\end{eqnarray}
The ratio of the deep inelastic structure functions can therefore be 
expressed as 
\begin{equation}
\frac{F_L(x,Q^2)}{F_2(x,Q^2)} = 2 \frac{1+(1-y)^2}{1-y 
+ 2y^2 A_{\cos 2\phi}^{(2)}(x,Q^2) } A_{\cos 2\phi}^{(2)}(x,Q^2)
\end{equation}

It should be noted that the sum over all hadrons $h$ in the above 
equations also includes
neutral hadrons, whose momentum direction is difficult to track
experimentally. Restricting the summation to charged hadrons only would 
introduce subprocess-dependent correction factors, relating to the
different probabilities of quarks and gluons to fragment into charged
hadrons, thus altering the relative magnitude of quark and gluon
contributions in the above asymmetry.  Furthermore, the summation
includes only hadrons that participated in the hard scattering process,
i.e.\ it excludes the proton remnant. Hadrons from the proton remnant
will however emerge with low $z$, their contribution in the second
moment can therefore be expected to be small. 

In summary, we have demonstrated that the longitudinal structure
function $F_L(x,Q^2)$ is related to the second moment of the 
angular $\cos 2\phi$ distribution of hadrons produced in
semi-inclusive deep inelastic scattering. The asymmetry defined in
(\ref{eq:master}) provides a possibility of measuring the structure
function ratio $F_L(x,Q^2)/F_2(x,Q^2)$ in a deep inelastic scattering
experiment at fixed collision energy. 

\section*{Acknowledgement}
The author would like to thank Jim Crittenden and 
Nick Brook for pointing his attention to
the issue discussed in this note and for numerous valuable discussions.

\end{document}